\documentclass[12pt]{article}
\usepackage[dvipdfmx]{graphicx}
\usepackage[utf8]{inputenc}
\usepackage{amsmath,amssymb,amscd}
\usepackage{ascmac}
\usepackage{latexsym}
\usepackage{bm}
\usepackage{booktabs}
\usepackage{mathrsfs}
\usepackage{float}
\usepackage{fancybox}
\usepackage{braket}
\usepackage{slashed}
\usepackage{float}
\usepackage{cases}
\usepackage{fancybox,pifont,layout}
\usepackage{amsmath,amsthm}

\usepackage{mediabb}
\usepackage{cite}
\usepackage{url}
\usepackage{arydshln}
\newcommand{\al}[1]{\begin{align}#1\end{align}} 

\usepackage[T1]{fontenc}
\usepackage{hyperref}
\hypersetup{ 
 setpagesize=false,
 bookmarksnumbered=true,%
 bookmarksopen=true,%
 colorlinks=true,%
 linkcolor=blue,
 citecolor=red,
}

\usepackage{color}
\usepackage[square,sort&compress,comma,numbers]{natbib}
\renewcommand{\thefootnote}{(\roman{footnote})}

\newcommand{\cred}{\textcolor{red}}

\makeatletter

\@addtoreset{equation}{section}
\makeatother

\begin{document}
\baselineskip=18pt
\begin{titlepage}
\begin{flushright}
{KOBE-TH-19-04}\\
\end{flushright}
\vspace{1cm}
\begin{center}{\Large\bf 
Dynamical generation of quark/lepton mass hierarchy in an extra dimension
}
\end{center}
\vspace{0.5cm}
\begin{center}
Yukihiro Fujimoto$^{(a)}$\footnote{E-mail: \url{y-fujimoto@oita-ct.ac.jp}},
Kouhei Hasegawa$^{(b)}$\footnote{E-mail: \url{kouhei@phys.sci.kobe-u.ac.jp}},
Kenji Nishiwaki$^{(c)}$\footnote{E-mail: \url{knishiw@irb.hr}},\\
Makoto Sakamoto$^{(b)}$\footnote{E-mail: \url{dragon@kobe-u.ac.jp}},
Kazunori Takenaga$^{(d)}$\footnote{E-mail: \url{takenaga@kumamoto-hsu.ac.jp}},
Pedro Hugo Tanaka$^{(b)}$\footnote{E-mail: {\url{186s116s@stu.kobe-u.ac.jp}}}\\
 and
Inori Ueba$^{(b)}$\footnote{E-mail: \url{i-ueba@stu.kobe-u.ac.jp}}
\end{center}
\begin{center}
${}^{(a)}${\it National Institute of Technology, Oita college, \\
Maki1666, Oaza, Oita 870-0152, Japan}\\
${}^{(b)}${\it Department of Physics, Kobe University, 
Kobe 657-8501, Japan}\\
${}^{(c)}${\it Ru{\dj}er Bo{\v{s}}kovi{\'c} Institute, Division of Theoretical Physics, Zagreb 10000, Croatia}\\
${}^{(d)}${\it {Faculty of Health Science}, Kumamoto Health Science University\\
325 Izumi-machi, Kita-ku, Kumamoto 861-5598, Japan }
\end{center}

\vspace{1cm}
\begin{abstract}
We show that the observed quark/lepton mass hierarchy can be realized 
dynamically on an interval extra dimension with point interactions. 
In our model, the positions of the point interactions play a crucial role to control the quark/lepton mass hierarchy and are determined by the minimization of the Casimir energy. 
By use of the exact extra-dimensional coordinate-dependent vacuum expectation value of a gauge singlet scalar, 
we find that there is a parameter set, where the positions of the point interactions are stabilized and fixed, 
which can reproduce the experimental values of the quark masses precisely enough, while the charged lepton part is less relevant.
We also show that possible mixings among the charged leptons will improve the situation significantly.
\end{abstract}
\end{titlepage}
\newpage

\setcounter{footnote}{0}
\renewcommand{\thefootnote}{\arabic{footnote}\,}
\section{Introduction}

{Even after} the discovery of the Higgs boson, {several issues associated with the Higgs boson} still remain 
to be solved in the Standard Model~(SM). 
One is {why the observed quark/lepton mass hierarchy is so drastic}. In the SM, quarks and leptons acquire their masses
through the Yukawa interactions {where they pick up} the vacuum expectation value~(VEV) of the Higgs field.
{Here, we should accept to impose ${\cal O}(10^{5})$-order hierarchical values on Yukawa couplings,
which look very unnatural.}

{Another unnatural request from the SM is associated with the matter content of the fermions}.
In both {of} the quark and the lepton sectors, {there are three similar but different copies} of quarks and leptons, 
which possess exactly the same {quantum numbers} except for their masses. 
{The origin of such generation structure is beyond the scope of the SM, and is still unveiled}.

One fascinating way to {address the common origin of the above two issues} is to introduce
{the objects so-called} point interactions on an interval extra dimension {in five dimensions~(5d)}~\footnote{Another way is to introduce the magnetic flux or the magnerized orbifold in the extra dimensions~\cite{Cremades:2004wa,Abe:2008fi,Abe:2008sx,Abe:2014noa,Abe:2014vza,Abe:2015yva,Fujimoto:2016zjs,Sakamura:2016kqv,Kobayashi:2016qag,Ishida:2017avx,Buchmuller:2017vho,Ishida:2018bbl,Abe:2018ylo,Abe:2018qbp}. There are also several ways to address the SM problems in the context of four-dimensional gauge theories, by use of noncompact gauge symmetry~\cite{Inoue:1994qz,Inoue:2000ia,Inoue:2003qi,Inoue:2007uj,Yamatsu:2012rj}.}. 
The point interactions provide {extra boundary conditions for each of 5d quarks and 5d leptons, and can induce} the degenerated {chiral} zero modes. {This is interpreted as spontaneous generation of the three matter generations in the SM
when we introduce two point interactions for each 5d fermion}. 
Appealing properties in this {direction} are that
{a 5d fermion leads to three chiral zero modes, and they are localized to different segments each other, 
whose end points are identified by the two corresponding point interactions}.
Due to {these properties, a three-by-three mass matrix is realized as $m_{ij} \,(i,j = 1,2,3)$}. 
{The generation dependence of the four-dimensional (4d) mass matrix $m_{ij}$ is described through the following manner:}
	\begin{align}
	{m_{ij} \propto \int_{0}^{L} dy}\,
	{\langle \Phi(y)\rangle} \bigl(g_{\psi_{i}{\rm{L}}}^{{(0)}}(y)\bigr)^{\ast}  f_{\psi'_{j}{\rm{R}}}^{{(0)}}(y),
	\label{overlap-integral}
	\end{align}
{where $L$ represents the length of the interval,}
{$\int^{L}_{0} dy$ is an integration along the shown range of the extra-dimensional coordinate}, 
$g_{\psi_{i}{\rm{L}}}^{{(0)}}(y)$ $\bigl(f_{\psi'_{j}{\rm{R}}}^{{(0)}}(y)\bigr)$ describes 
{mode functions} of {\it i}-th ({\it j}-th) generation left- (right-) handed {chiral} zero mode $\psi_{i{\rm{L}}}(x)$ $\bigl(\psi'_{j{\rm{R}}}(x)\bigr)$.
{$\langle \Phi(y)\rangle$ represents the contribution from the classical configuration of the scalar which appears
in the 5d Yukawa term and its VEV shows the dependence on $y$.}
{If not only $g_{\psi_{i}{\rm{L}}}^{{(0)}}(y)$ and $f_{\psi'_{j}{\rm{R}}}^{{(0)}}(y)$, but also $\langle \Phi(y) \rangle$} are localized functions,
{sizable mass hierarchy can appear through the overlap integrals in Eq.}~(\ref{overlap-integral}).

In this paper, based on the above mechanism and the previous researches \cite{Fujimoto:2012wv,Fujimoto:2014fka,Fujimoto:2013ki}, we perform a numerical research to {appraise whether the scenario can reproduce the observed values of the quark/lepton hierarchical masses dynamically} (see~\cite{Panico:2016ull,Agashe:2016rle,DaRold:2017xdm,Ahmed:2019zxm} for dynamical
generations of the mass hierarchy). The positions of the point interactions, which play an important role to control the magnitude of the overlap integrals, seem to be free parameters. {Nonetheless, they are not free but are} determined dynamically {through} the vacuum configuration to minimize the Casimir energy
(see~\cite{Ponton:2001hq,deAlbuquerque:2003qbk} and
also~\cite{Garriga:2000jb,Goldberger:2000dv,Hofmann:2000cj,Abe:2014eia}
as related works). 
{This dynamical determination of the positions of the point interactions} reduces the number of the free parameters.
{Also, to accelerate hierarchical values in} the overlap integral in Eq.~(\ref{overlap-integral}),
we {employ} {a new type of extra-dimensional coordinate-dependent} {configuration in $\langle \Phi(y)\rangle$},
which is also obtained dynamically by {solving the equation of motion with the quartic interaction}. 
It should be {noted} that in the prior research \cite{Fujimoto:2017lln}, {the VEV of the scalar $\langle \Phi(y)\rangle$ is also obtained dynamically.
{However,} the different functional type of the VEV was adopted} {(see subsection \ref{sec:classical-Phi-details} in detail)}.
{Performing numerical searches brings us the conclusion that the hierarchical masses of the quarks are fully reproduced,
while the charged leptons do not work well.
Introduction of flavor mixings may improve our current results, especially for the charged leptons.}

The paper is organized as follows.
In section~\ref{sec:section-2}, we briefly review the extra-dimensional model with point interactions on an interval. 
{In section~\ref{sec:section-3}, we discuss the dynamics of the Casimir energy, the potential minimization of the scalar $\Phi$ and
their implication for the mass hierarchy}. 
In section~\ref{sec:section-4}, we {investigate searches for sets of desirable input parameters} and show our results. 
Section~\ref{sec:section-5} is devoted to conclusion and discussion.

\section{{The} model with point interactions on an interval
\label{sec:section-2}}

\subsection{Basic idea}

\begin{figure}[ht]
		\begin{center}
		\includegraphics[width=100mm]{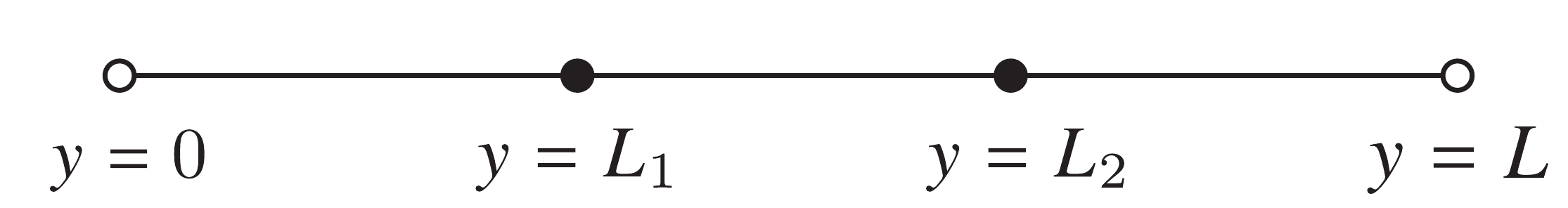}
		\par
		\vspace{0.0cm}
		\caption{A schematic figure of the interval extra dimension with the point interactions. The black dots at $y=L_{1},L_{2}$ indicate the point interaction.}
		\label{fig:extra_dim-PIs}
		\end{center}
		\begin{center}
		\includegraphics[width=170mm]{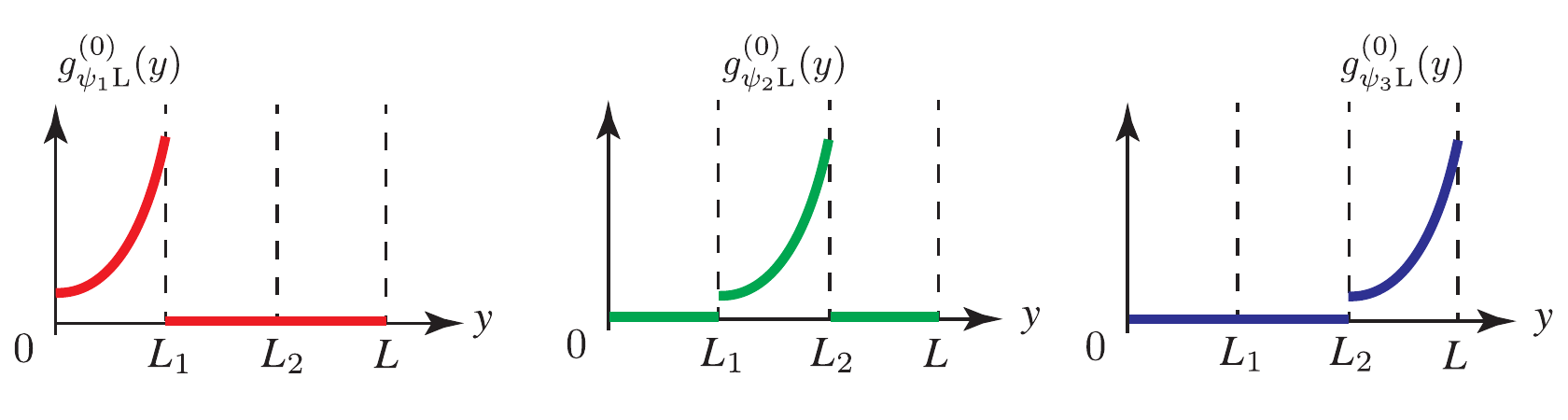}
		\par
		\vspace{0.0cm}
		\caption{A schematic figure of the profiles of the three {chiral} zero modes. 
		Each zero mode localizes towards {the different boundary points but in a similar way}.}
		\label{fig:G_0}
		\end{center}
\end{figure}

In this subsection, we provide a brief review on {the extra-dimensional model} which sheds light on spontaneous generation of
the three generations of the fermion matter content, the hierarchical structure of the observed quarks and charged leptons,
and their mixing structures in the SM, simultaneously.
This kind of scenario was firstly proposed in~\cite{Fujimoto:2012wv} on an interval to explore the quark sector, where
objects so-called point interactions play a significant role.
The point interactions describe possible singularities in one-dimensional quantum mechanics, by which profiles of particles
along the extra spacial direction $y$ are represented.
This statement is rephrased that we can add `extra (or generalized) boundary conditions' for all kinds of 5d fields in the bulk space,
in addition to the boundary points (see Appendix~A of~\cite{Fujimoto:2017lln} for more details).
{{Three} {chiral} zero modes are generated} from a single 5d fermion by imposing {the Dirichlet boundary conditions at the point interactions}~\cite{Fujimoto:2012wv}.
A key point to realize mass hierarchy is that we can introduce a bulk mass for each of such 5d fermions {respectively}, which
makes {the profiles of the three zero modes} localized towards the boundary points {as shown in Fig.~\ref{fig:extra_dim-PIs} and Fig.~\ref{fig:G_0}}.

Here, if the scalar appearing in the Yukawa sector of such 5d fermions possesses {an extra-dimensional coordinate-dependent VEV},
it is clear that sizable mass hierarchy is generated on an interval.
It is noted that this scenario has been also developed in the $S^1$ geometry, where a complex phase for describing 
the {{CP}-violating phase} of the {Cabbibo--Kobayashi--Maskawa (CKM)} matrix can emerge, originating from a twisted boundary condition~\cite{Fujimoto:2013ki,Fujimoto:2014fka}.

In the scenario, an $SU(3)_C \times SU(2)_L \times U(1)_Y$ gauge theory is unfolded on an interval, the relevant part of the 5d action
of which for our discussion reads,
\al{
S &= S_\text{quark} + S_\text{lepton} + S_\text{Higgs} + S_\text{singlet} + S_\text{Yukawa}, \\[5pt]
S_\text{quark}
&=
\int \!\! d^4x \!\! \int_0^L \!\! dy 
\left[ 
\overline{Q} \left( i \Gamma^M D^{{(Q)}}_M + M_Q \right) Q + 
\overline{\mathcal{U}} \left( i \Gamma^M D^{{(\mathcal{U})}}_M + M_\mathcal{U} \right) \mathcal{U} +
\overline{{\cal D}} \left( i \Gamma^M D^{{({\cal D})}}_M + M_\mathcal{D} \right) \mathcal{D} 
\right], \\
S_\text{lepton}
&=
\int \!\! d^4x \!\! \int_0^L \!\! dy 
\left[ 
\overline{L} \left( i \Gamma^M D^{{(L)}}_M + M_L \right) L + 
\overline{\mathcal{E}} \left( i \Gamma^M D^{{(\mathcal{E})}}_M + M_\mathcal{E} \right) \mathcal{E}
\right], \\
S_\text{Higgs}
&=
\int \!\! d^4x \!\! \int_0^L \!\! dy 
\left[ 
H^\dagger \left( D^M D_M + M^2_{H} \right) H - \frac{\lambda_H}{2} (H^\dagger H)^2
\right], \\
S_\text{singlet}
&=
\int \!\! d^4x \!\! \int_0^L \!\! dy 
\left[ 
\Phi^\dagger \left( \partial^M \partial_M - M^2_{\Phi} \right) \Phi - \frac{\lambda_\Phi}{2} (\Phi^\dagger \Phi)^2
\right], 
	\label{eq:S_singlet} \\
S_\text{Yukawa}
&=
\int \!\! d^4x \!\! \int_0^L \!\! dy 
\left[ 
\Phi \left(  
- \mathcal{Y}_{u} \overline{Q} (i \sigma_2 H^\ast) \mathcal{U}
- \mathcal{Y}_{d} \overline{Q} H \mathcal{D}
- \mathcal{Y}_{e} \overline{L} H \mathcal{E}  
\right) + (\text{h.c.})
\right],
}
which consists of a 5d quark doublet field $Q$ and a 5d lepton doublet field $L$;
singlet fields for an up-type quark $\mathcal{U}$, a down-type quark $\mathcal{D}$ and a charged lepton $\mathcal{E}$;
the Higgs doublet $H$ and a gauge singlet complex scalar $\Phi$; the Pauli matrix $\sigma_{2}$, where
we skipped to show the apparent gauge sector of the current scenario and
we usually suppress the 5d coordinates as variables of 5d fields.
In this manuscript, we do not discuss the minuscule active neutrino mass generation {so that} the (5d) neutrino singlet field
is not introduced (see the discussions on the $S^1$ geometry without/with Majorana-like mass terms~\cite{Fujimoto:2014fka}/\cite{Cai:2015jla}).
The capital latin alphabets $M, N, \cdots$ represent the 5d Minkowski indices, where the 4d part is described by {the Greek characters} $\mu, \nu, \cdots$.
The 5d coordinate $x^M$ is thus decomposed as $\{ x^\mu, y \}$ {where $x^{\mu}$ denotes the 4d Minkowski coordinate and $y$ denotes the extra-dimensional one}.
Our convention for the 5d flat metric reads $\eta_{MN} = \text{diag}(-1,+1,+1,+1,+1)$ and the representation of the 5d Clifford 
algebra is chosen as $\{ \Gamma_M, \Gamma_N \} = -2 \eta_{MN}$, where the concrete forms of the 5d gamma matrices
are adopted as $\Gamma_\mu = \gamma_\mu$ and $\Gamma_y = -i \gamma_5 = \gamma^0 \gamma^1 \gamma^2 \gamma^3$.
Various $D^{{(\Psi)}}_M$ {$(\Psi=Q,\mathcal{U},{\cal D},L,\mathcal{E})$} symbols represent the corresponding covariant derivatives,
{$M_{\Psi}$ $(\Psi=Q, {\cal U}, {\cal D}, L, {\cal E})$} are bulk masses for the 5d fermions,
{$M^2_{H}$ ($M^{2}_{\Phi}$) and $\lambda_{H}$ ($\lambda_{\Phi}$) are squared mass scales and quartic couplings for the Higgs (the gauge-singlet scalar)}{. $\mathcal{Y}_{u},\mathcal{Y}_{d}$ and $\mathcal{Y}_{e}$} are overall Yukawa couplings which do not hold indices to discriminate matter generations {since all the 5d fermions {are assumed to be one-generation}}.
The sign in front of  $M_H^2$ is chosen to ignite the Higgs mechanism.
Obedience to the conditions {$\lambda_{H}>0$ and $\lambda_{\Phi}>0$} is compulsory to ensure the stability of the whole potential.\footnote{
In general, the mixing term $(H^\dagger H)(\Phi^\dagger \Phi)$ {can be written down. However,} the existence of this term leads to gauge universality violation~\cite{Fujimoto:2012wv}{. Then}
we set the coefficient as zero by hand.
}

Here we provide several comments.
\begin{itemize}
\item
As we mentioned above, we introduce two point interactions for each of the 5d fermions to realize {three generations at the {chiral} zero mode sector},
while not for the scalars and the gauge bosons.
\item
We impose the discrete symmetry, $H \to - H$ and $\Phi \to - \Phi$, which
prohibits the `ordinary' Yukawa terms like $\overline{Q}(i \sigma_2 H^\ast) {\cal U}$ and $\Phi \overline{Q} Q$ and others likewise.
We assume that $H$ has the ordinary {constant VEV $\langle H\rangle$} to break the SM gauge group suitably, and
$\Phi$ acquires a $y$-dependent VEV {$\langle \Phi\rangle$} to generate hierarchical structure in the Yukawa sector.
Thus, the above `higher-dimensional' Yukawa terms are suitable for our purpose.
The detail of the {$y$-dependent gauge-singlet scalar VEV profile will be provided in subsection~\ref{sec:classical-Phi-details}.} 
\item
Under the current setup on an interval, no physical CP-violating phase can emerge since only overall complexities can be realized
by {the Yukawa couplings $\mathcal{Y}_{u},\mathcal{Y}_{d}$ and $\mathcal{Y}_{e}$} and absorbed into field redefinitions.
We conquer this obstacle by moving to the $S^1$ geometry easily~\cite{Fujimoto:2013ki,Fujimoto:2014fka}, 
while we do not discuss this point in this paper.
\end{itemize}

\subsection{{Appearance of {three-generation mass matrices}}}

As discussed in~\cite{Fujimoto:2012wv},
introduction of two suitable Dirichlet boundary conditions by point interactions for a 5d fermion $\Psi$ in the bulk space,
in addition to the two endpoints of the interval, as $\Psi_{\rm{R}} \equiv \left( \frac{1+\gamma_5}{2} \right) \Psi = 0\quad \text{or} \quad
\Psi_{\rm{L}} \equiv \left( \frac{1-\gamma_5}{2} \right) \Psi = 0$ leads to the realization of three localized left- or right-handed {{chiral}} zero modes via $\Psi$, respectively.
Concretely, the following conditions are imposed for the 5d {fermions} with an infinitesimal positive constant $\varepsilon$:
\al{
Q_{\rm{R}} &= 0 \ \ \ \ \ \ \ \  \text{at} \ \ y 
= {0}, \, L_1^{(Q)} \pm \varepsilon, \, L_2^{(Q)} \pm \varepsilon, \, {L}, \\
\mathcal{U}_{\rm{L}} &= 0  \ \ \ \ \ \ \ \ \text{at} \ \ y 
= {0}, \, L_1^{(\mathcal{U})} \pm \varepsilon, \, L_2^{(\mathcal{U})} \pm \varepsilon, \,{L}, \\
\mathcal{D}_{\rm{L}} &= 0  \ \ \ \ \ \ \ \ \text{at} \ \ y 
= {0}, \, L_1^{(\mathcal{D})} \pm \varepsilon, \, L_2^{(\mathcal{D})} \pm \varepsilon, \,{L}, \\
L_{\rm{R}} &= 0  \ \ \ \ \ \ \ \  \text{at} \ \ y 
={0}, \, L_1^{(L)} \pm \varepsilon, \, L_2^{(L)} \pm \varepsilon, \, {L}, \\
\mathcal{E}_{\rm{L}} &= 0  \ \ \ \ \ \ \ \  \text{at} \ \ y 
= {0}, \, L_1^{(\mathcal{E})} \pm \varepsilon, \, L_2^{(\mathcal{E})} \pm \varepsilon, \, {L}, 
}
where {$L_{1}^{(\Psi)}$ and $L_{2}^{(\Psi)}$} mean the positions of the first and second point interactions ($0 < L_{1}^{(\Psi)} < L_{2}^{(\Psi)} < L$) for the field $\Psi$ {$(\Psi=Q,\mathcal{U},{\cal D},L,\mathcal{E})$}.
Here we define the conventions $L_{0}^{(\Psi)} \equiv 0$ and $L_{3}^{(\Psi)} \equiv L$ for convenience soon later.
The Kaluza--Klein~(KK) expansions of the 5d fermions are represented as
\al{
Q(x,y) = \begin{pmatrix} U(x,y) \\ D(x,y) \end{pmatrix}
&=  {\sum_{i=1}^{3}\begin{pmatrix} u_{i{\rm{L}}}^{(0)}(x)  \\  d_{i{\rm{L}}}^{(0)}(x)  \end{pmatrix}g^{(0)}_{q_{i{\rm{L}}}}(y)
}+ (\text{massive modes}), \\
\mathcal{U}(x,y)
&= \sum_{i=1}^{3} u_{i{\rm{R}}}^{(0)}(x) {f^{(0)}_{u_{i{\rm{R}}}}(y)}
+ (\text{massive modes}), \\
\mathcal{D}(x,y)
&= \sum_{i=1}^{3} d_{i{\rm{R}}}^{(0)}(x) {f^{(0)}_{d_{i{\rm{R}}}}(y)}
+ (\text{massive modes}), \\
L(x,y) = \begin{pmatrix} N(x,y) \\ E(x,y) \end{pmatrix}
&={ \sum_{i=1}^{3} \begin{pmatrix} \nu_{i{\rm{L}}}^{(0)}(x) \\ \ e_{i{\rm{L}}}^{(0)}(x)  \end{pmatrix}
g^{(0)}_{l_{i{\rm{L}}}}(y) }+ (\text{massive modes}), \\
\mathcal{E}(x,y)
&= \sum_{i=1}^{3} e_{i{\rm{R}}}^{(0)}(x) {f^{(0)}_{e_{i{\rm{R}}}}(y)}
+ (\text{massive modes}),
} 
with the generation indices $i\ (= 1, 2, 3)$.
The concrete forms of the zero-mode profiles are easily derived as~\cite{Fujimoto:2012wv}
\al{
{g^{(0)}_{q_{i{\rm{L}}}}(y)}
&=
{\bigl[ \theta(y - L^{(Q)}_{i-1}) \, \theta(L^{(Q)}_{i} - y) \bigr]}\cdot \mathcal{N}_i^{(Q)} e^{+ M_Q (y - L^{(Q)}_{i-1})} 
, \\
{f^{(0)}_{u_{i{\rm{R}}}}(y)}
&={\bigl[ \theta(y - L^{(\mathcal{U})}_{i-1})\theta(L^{(\mathcal{U})}_{i} - y) \bigr]}\cdot
\mathcal{N}_i^{(\mathcal{U})} e^{- M_\mathcal{U} (y - L^{(\mathcal{U})}_{i-1})}
 , \\
{f^{(0)}_{d_{i{\rm{R}}}}(y)}
&={\bigl[ \theta(y - L^{(\mathcal{D})}_{i-1}) \, \theta(L^{(\mathcal{D})}_{i} - y) \bigr]}\cdot
\mathcal{N}_i^{(\mathcal{D})} e^{- M_\mathcal{D} (y - L^{(\mathcal{D})}_{i-1})}, \\
{g^{(0)}_{\cred{l}_{i{\rm{L}}}}(y)}
&={\bigl[ \theta(y - L^{(L)}_{i-1}) \, \theta(L^{(L)}_{i} - y) \bigr]}\cdot
\mathcal{N}_i^{(L)} e^{+ M_L (y - L^{(L)}_{i-1})}, \\
{f^{(0)}_{e_{i{\rm{R}}}}(y)}
&={\bigl[ \theta(y - L^{(\mathcal{E})}_{i-1}) \, \theta(L^{(\mathcal{E})}_{i} - y) \bigr]}\cdot 
\mathcal{N}_i^{(\mathcal{E})} e^{- M_\mathcal{E} (y - L^{(\mathcal{E})}_{i-1})},\ \ (i=1, 2, 3) 
}
where {$\theta (y)$ denotes the Heaviside step function} and the individual kinetic normalization constants are expressed
\al{
\mathcal{N}_i^{(Q)} &= \sqrt{ \frac{ 2 M_Q }{ e^{2 M_Q \Delta L_i^{(Q)}} -1 } } \,, \qquad
\mathcal{N}_i^{(\mathcal{U})} = \sqrt{ \frac{ 2 M_\mathcal{U} }{ 1 - e^{ - 2 M_\mathcal{U} \Delta L_i^{(\mathcal{U})}} } } \,, \qquad
\mathcal{N}_i^{(\mathcal{D})} = \sqrt{ \frac{ 2 M_\mathcal{D} }{ 1 - e^{ - 2 M_\mathcal{D} \Delta L_i^{(\mathcal{D})}} } } \,, \notag \\
\mathcal{N}_i^{(L)} &= \sqrt{ \frac{ 2 M_L }{ e^{2 M_L \Delta L_i^{(L)}} -1 } } \,, \qquad
\mathcal{N}_i^{(\mathcal{E})} = \sqrt{ \frac{ 2 M_\mathcal{E} }{ 1 - e^{ - 2 M_\mathcal{E} \Delta L_i^{(\mathcal{E})}} } } \,, \ \ \ \ \ \ (i=1, 2, 3) 
}
with the lengths of segments
\al{
\Delta L_i^{(\Psi)} \equiv L_i^{(\Psi)} - L_{i-1}^{(\Psi)} \qquad
\left( \text{for} \ \ i = 1,2,3; \Psi = Q, \mathcal{U}, \mathcal{D}, L, \mathcal{E} \right).
}

After the integration along the $y$ direction, we obtain three-by-three mass matrices, {\it e.g.,} for the up-type quarks as
\al{
S_\text{up-type quark mass}\ =
- {\int d^{4}x}\,\sum_{i,j = 1}^{3} \overline{u}^{(0)}_{i {\rm{L}}} M^{(u)}_{ij} u^{(0)}_{j{\rm{R}}} + (\text{h.c.}).
}
It should be emphasized that the ordering of the positions of point interactions governs the form of the mass matrices.
For example, when the following ordering is realized{,}
\al{
0 < L_1^{(\cal U)} < L_1^{(Q)} < L_2^{(\cal U)} < L_2^{(Q)} < L,
          \label{less:L-conditions}
}
the corresponding concrete forms of the elements of the up-type quark mass matrix are taken
\al{
M^{(u)}
=
\begin{bmatrix}
M^{(u)}_{11} & M^{(u)}_{12} & 0 \\
0 & M^{(u)}_{22} & M^{(u)}_{23} \\
0 & 0 & M^{(u)}_{33}
\end{bmatrix},
	\label{eq:Mu-example}
}
with
\al{
M^{(u)}_{11} \
&= \frac{\mathcal{Y}_{u} {v}}{\sqrt{2}} { \int_0^{L_1^{(\cal U)}} \!\!\!\! dy } \ \langle \Phi(y) \rangle {g^{(0)}_{q_{1{\rm{L}}}}(y) f^{(0)}_{u_{1{\rm{R}}}}(y)}, 
	\label{eq:Mu-first} \\
M^{(u)}_{22} \
&= \frac{\mathcal{Y}_{u} {v}}{\sqrt{2}} { \int_{L_1^{(Q)}}^{L_2^{(\cal U)}} \!\!\!\! dy } \ \langle \Phi(y) \rangle {g^{(0)}_{q_{2{\rm{L}}}}(y) f^{(0)}_{u_{2{\rm{R}}}}(y)}, \\
M^{(u)}_{33} \
&= \frac{\mathcal{Y}_{u} {v}}{\sqrt{2}} { \int_{L_2^{(Q)}}^{L_3^{(\cal U)}} \!\!\!\! dy } \ \langle \Phi(y) \rangle {g^{(0)}_{q_{3{\rm{L}}}}(y) f^{(0)}_{u_{3{\rm{R}}}}(y)}, \\
M^{(u)}_{12} \
&= \frac{\mathcal{Y}_{u} {v}}{\sqrt{2}} { \int_{L_1^{(\cal U)}}^{L_1^{(Q)}} \!\!\!\! dy } \ \langle \Phi(y) \rangle {g^{(0)}_{q_{1{\rm{L}}}}(y) f^{(0)}_{u_{2{\rm{R}}}}(y)}, \\
M^{(u)}_{23} \
&= \frac{\mathcal{Y}_{u} {v}}{\sqrt{2}} { \int_{L_2^{(\cal U)}}^{L_2^{(Q)}} \!\!\!\! dy } \ \langle \Phi(y) \rangle {g^{(0)}_{q_{2{\rm{L}}}}(y) f^{(0)}_{u_{3{\rm{R}}}}(y)}, 
	\label{eq:Mu-last}
}
where we assumed the form for the Higgs doublet, $\langle H \rangle = (0, {v}/\sqrt{2})^\text{T}$,
{where $v$ is the 5d Higgs VEV and can be treated as real without loss of generality as in the SM}.
If the ordering is different from Eq.~(\ref{less:L-conditions}), we can easily obtain the corresponding forms by looking at the profiles of fermion mode functions.
See~\cite{Fujimoto:2012wv} (also~\cite{Fujimoto:2013ki,Fujimoto:2014fka}) for more details.

\section{{Dynamics in the scenario} {and their implications}
\label{sec:section-3}}

We first mention that, in~\cite{Fujimoto:2012wv}, the observed quark mass hierarchy and the three mixing angles of the CKM matrix
were successfully reproduced through the current strategy, under the following {preconditions}:
\begin{itemize}
\item
All of the positions of the point interactions can be treated as individual free parameters.
\item
{The form of the $\langle \Phi(y) \rangle$ is exponential as $\langle \Phi(y) \rangle \sim e^{M_\Phi y}$ (where the mass scale is not shown),
whose form can be materialized by setting the parameters associated with $\Phi$ appropriately
in the actual form of $\langle \Phi(y) \rangle$ (in Case (II) in subsection~\ref{sec:classical-Phi-details})~\cite{Fujimoto:2012wv}}.
\end{itemize}
These {preconditions} would be justified for phenomenological studies.
However, the following criticisms will come in various points of view:
\begin{itemize}
\item[(a)] {\bf From stability of the system}:
	the positions of the point interactions contribute to the background Casimir energy of this extra-dimensional system, which should be
	minimized (or extremized at least) to ensure the stability of the whole system.
	If one obeys this line, the positions of the point interactions cannot take arbitrary values.
\item[(b)] {\bf From the number of free parameters}:
	for quarks, in addition to the six parameters as the positions of point interactions for quark fields
	{\{$L_{i}^{(Q)}, L_{i}^{(\mathcal{U})}, L_{i}^{(\mathcal{D})};i=1,2\}$},
	the following parameters join to describe the quark profiles naively:
	one as the length of the system ${L}$,
	two as the overall Yukawa couplings ${\{\mathcal{Y}_{u},\mathcal{Y}_{d}\}}$,
	three as the bulk masses ${\{M_{Q}, M_{\mathcal{U}}, M_{\mathcal{D}}\}}$,
	one as the 5d Higgs VEV ${v}$,
	and four to {the gauge-singlet scalar VEV $\langle\Phi(y)\rangle$} (where details are provided in subsection~\ref{sec:classical-Phi-details});
	there are $17$ free parameters in total (for the {lepton} sector, where apparently six additional ones 
	{\{$L_{1}^{(L)}$, $L_{2}^{(L)}$, $L_{1}^{(\mathcal{E})}$, $L_{2}^{(\mathcal{E})}$, $M_{L}$, $M_{\mathcal{E}}$\}} join when we take care of charged leptons).\footnote{
	As we will see in Section~\ref{sec:diagonal-numerics}, this counting of parameters is naive a bit.
	The number of the degrees of freedom relevant for mass ratios is less than the digit.
	}
	On the contrary, only the nine digits (six for mass eigenvalues, three for CKM mixing angles)
	are fitted.
	Such an unbalanced situation may lead to the criticism that the result of the previous fit is not surprising, looks trivial.
\item[(c)] {\bf From generality in the profile of $\langle \Phi(y) \rangle$}:
	As partially discussed in Section~4 of~\cite{Fujimoto:2012wv}, the actual form of $\langle \Phi(y) \rangle$ is {provided by the Jacobi's elliptic function. As we will see in sebsection~\ref{sec:classical-Phi-details}, there are two solutions for the VEV $\langle \Phi(y) \rangle$, which may give} new insights into the geometric realization of the observed Yukawa texture in the `generalized' setup.
\end{itemize}

Based on the result of~\cite{Fujimoto:2017lln} and the general information of $\langle \Phi(y) \rangle$,
we can provide our answers for all of the criticisms, at least partially.

\subsection{{Criterion via stabilization of point interactions}}

{The determination of the positions of the point interactions $L_i^{(\Psi)}$ by the minimization of the Casimir energy has been developed in~\cite{Fujimoto:2017lln}. The Casimir energy $E$ at one-loop order as the function of $L_i^{(\Psi)}$ takes to be the form of}
	\al{
	E=2\sum_{\Psi=Q,L}E^{(\Psi)}[L^{(\Psi)}_{1},L^{(\Psi)}_{2}]+\sum_{\Psi'={\cal U},{\cal D},{\cal E}}E^{(\Psi')}[L^{(\Psi')}_{1},L^{(\Psi')}_{2}],	
	}
where
	\al{
	E^{(\Psi)}[L^{(\Psi)}_{1},L^{(\Psi)}_{2}]&=\sum_{i=1}^{3}\frac{|M^{(\Psi)}|^{2}}{8\pi^{2}\bigl(L_{i}^{(\Psi)}-L_{i-1}^{(\Psi)}\bigr)^{2}}\sum^{\infty}_{w=1}\frac{e^{-2w|M^{(\Psi)}|(L_{i}^{(\Psi)}-L_{i-1}^{(\Psi)})}}{w^{3}}\nonumber\\
	&\qquad \times\left(1+\frac{3}{2w|M^{(\Psi)}|\bigl(L_{i}^{(\Psi)}-L_{i-1}^{(\Psi)}\bigr)}+\frac{3}{4w^{2}|M^{(\Psi)}|^{2}\bigl(L_{i}^{(\Psi)}-L_{i-1}^{(\Psi)}\bigr)^{2}}\right).
	}

{The minimization of the above Casimir energy leads to the results in which the positions of the point interactions for the doublet fermions and the singlet fermions are the same as \footnote{
{In the case of the discussion about the total length $L$ by the Casimir energy, the other fields contributions, {\it e.g.} gauge fields and the scalar fields, should be included to the minimization condition (see the discussion in~\cite{Fujimoto:2017lln}). On the other hand, in the case of the discussion about the positions of the point interactions, only the related fermion's contributions are important.}
}}
\al{
L_1^{(Q)} &= L_1^{(\mathcal{U})} = L_1^{(\mathcal{D})} {= L_1^{(L)} = L_1^{(\mathcal{E})}}, \notag \\
L_2^{(Q)} &= L_2^{(\mathcal{U})} = L_2^{(\mathcal{D})} {= L_2^{(L)} = L_2^{(\mathcal{E})}},
	\label{eq:MS-Ansatz}
}
{and all of the point interactions should be located at}
\al{
L_1^{(\Psi)} = \frac{1}{3} L, \qquad
L_{{2}}^{(\Psi)} = \frac{2}{3} L,
}
{to minimize the Casimir energy in spite of the type of the field $\Psi$. More concretely,}
\al{
L_1^{(Q)} = L_1^{(\mathcal{U})} = L_1^{(\mathcal{D})} = L_1^{(L)} = L_1^{(\mathcal{E})} &= \frac{1}{3} L, \notag \\
L_2^{(Q)} = L_2^{(\mathcal{U})} = L_2^{(\mathcal{D})} = L_2^{(L)} = L_2^{(\mathcal{E})} &= \frac{2}{3} L.
	\label{eq:condition-on-Ls}
}
It is noted that (\ref{eq:condition-on-Ls}) implies no mixing terms in mass matrices. Taking account of this result in parameter fitting would definitely become a reasonable response (despite of being not complete)
to the above criticisms (a) and (b), where the naive counting of the input parameters for quarks reduced to $11$, from $17$.

\subsection{Criterion via the general solution of {$\langle \Phi(y) \rangle$}
\label{sec:classical-Phi-details}}

To know the actual solution of $\Phi $ means to solve the following nonlinear equation derived
through the variational principle from $S_\text{singlet}$ in Eq.~(\ref{eq:S_singlet}),
\al{
\frac{d^2 \phi(y)}{d y^2} - M_\Phi^2 \phi(y) - \lambda_\Phi \phi^3(y) = 0, 
}
where we put the ansatz that the VEV of $\Phi$ depends on the extra dimensional coordinate $y$, {\it i.e.} 
\al{
{\langle \Phi \rangle = \phi(y)}\cred{,}
}
where in general the following form comes
$\langle \Phi \rangle = e^{i \theta(y)} \phi(y)$, but
we have proven that without loss of generality we can set the
$y$-dependent phase part $\theta(y)$ to be zero,
where $\phi(y)$ is real. The above equation leads to
\al{
\frac{d}{d y} \left[ \frac{1}{2} \left( \frac{d \phi(y)}{d y} \right)^2 - \frac{M_\Phi^2}{2} \phi^2(y) - \frac{\lambda_\Phi}{4} \phi^4(y)  \right] = 0,
}
and can be deformed to
\al{
\frac{1}{2} \left( \frac{d \phi(y)}{d y} \right)^2 + U(\phi) = E_\Phi, \qquad 
\left(  U(\phi) \equiv -\frac{M_\Phi^2}{2} \phi^2(y) - \frac{\lambda_\Phi}{4} \phi^4(y) \right),
}
with an undetermined integration constant $E_\Phi$ as a parameter.
Here, we should take account of the boundary conditions of $\phi(y)$ at $y=0$ and $y=L$, where
we impose the Robin boundary conditions, which are described with the length parameters $L_\pm$ {$(-\infty\leq L_{\pm}\leq +\infty)$} as below\footnote{
We note that the scalar field $\Phi$ is assumed not to feel point interactions.}
\al{
\phi(0) + L_+ \phi'(0) &= 0, \notag \\
\phi(L) - L_- \phi'(L) &= 0,
	\label{eq:Robin-BC}
}
where the prime symbol represents the derivative of $y$ (refer to~\cite{Fujimoto:2011kf}).

Through elliptic integrals, at least the following {\it two} solutions are possible in terms of Jacobi's elliptic functions:
\begin{itemize}
\item[(I)] When $0 < E_\Phi < M^4_\Phi/(4 \lambda_\Phi)$:{
\al{
\phi(y) = \mu_- \frac{\text{sn}\left( \mu_+ \sqrt{\frac{\lambda_\Phi}{2}} (y - y_0), k \right)}{\text{cn}\left( \mu_+ \sqrt{\frac{\lambda_\Phi}{2}} (y - y_0), k \right)},
}
where
\al{
\mu_\pm^{2} \equiv \frac{M^2_\Phi}{\lambda_\Phi} \left( 1 \pm \sqrt{1 - \frac{4 \lambda_\Phi E_{\Phi}}{M^4_\Phi} } \right),\quad
k^2 \equiv \frac{\mu^2_+ - \mu^2_-}{\mu^2_+}.
}
\item[(II)] When $E_\Phi < 0$:
\al{
\phi(y) = \frac{\nu}{\text{cn}\left( \frac{\mu}{k} \sqrt{\frac{\lambda_\Phi}{2}} (y - y_0), k \right)},
}
where
\al{
\mu^2 \equiv \frac{M^2_\Phi}{\lambda_\Phi} \left( 1 + \sqrt{1 - \frac{4 \lambda_\Phi E_{\Phi}}{M^4_\Phi} } \right),\quad
\nu^2 \equiv \frac{M^2_\Phi}{\lambda_\Phi} \left( \sqrt{1 - \frac{4 \lambda_\Phi E_{\Phi}}{M^4_\Phi} } - 1 \right),\quad
k^2 \equiv \frac{\mu^2}{\mu^2 + \nu^2}.
}}
\end{itemize}
We point out that the variables $L_\pm$ in Eq.~(\ref{eq:Robin-BC}) to parametrize the Robin boundary conditions are automatically fixed by the above profiles.
$y_0$ represents the degree of freedom of the translation along the $y$ direction.
The parameter $k$ measures ellipticity defined in the range of $0 \leq k \leq 1$,\footnote{
Actually, the limited region $1/\sqrt{2} \leq k \leq 1$ can be taken both in the cases of (I) and (II).
}
where the two extremal cases $k = 0$ and $1$ correspond to
\al{
\frac{\text{sn}(x,k)}{\text{cn}(x,k)} = \text{sc}(x,k) &\to
\begin{cases}
\tan{x} & \text{when} \ \ k \to 0, \\
\sinh{x} & \text{when} \ \ k \to 1,
\end{cases}
\notag \\
\frac{1}{\text{cn}(x,k)} &\to
\begin{cases}
1/\cos{x} & \text{when} \ \ k \to 0, \\
\cosh{x} & \text{when} \ \ k \to 1.
\end{cases}
}
When $k \not= 1$, a very nontrivial feature of the two elliptic functions $\text{sc}(x,k)$ and $1/\text{cn}(x,k)$ rises up,
where the functions get divergent periodically along the $y$-direction.
The minimal periods between the nearest divergent points of $\phi(y)$ in Case (I) and Case (II) are
\begin{equation*}
2\left( \mu_+ \sqrt{\frac{\lambda_\Phi}{2}} \right)^{-1} K[k],\quad
2\left( \frac{\mu}{k} \sqrt{\frac{\lambda_\Phi}{2}} \right)^{-1} K[k],
\end{equation*}
respectively with the complete elliptic integral of the first kind $K[k]$.
A possible important difference between Case (I) and Case (II) is that
$\text{sc}(x,k)$ increases monotonically from a divergent point $x_d$ $\bigl(\text{sc}(x_d + \varepsilon,k) \sim - \infty\bigr)$
to the next divergent point $x'_d \, (> x_d)$
$\bigl(\text{sc}(x'_d - \varepsilon,k) \sim + \infty\bigr)$, where the value of $\text{sc}(x,k)$ is negative in the first half part of the region {$x_d < x < (x_d + x'_d)/2$}.
This property would be useful to realize gigantic hierarchy in the overlap integrals that 
express the components of the {mass} matrices (see Eqs.(\ref{eq:Mu-first})--(\ref{eq:Mu-last}))
since negative contributions get to be possible.
This motivates us to have a discussion for exploring the situation with the exact form of $\phi(y)$ in Case (I) {in contrast with the prior case (II) researches. This will be the response to the criticism (c) shown above.}

\section{Numerical analysis {on an interval}
\label{sec:section-4}}

Following the discussions in the previous section,
we {perform} a {numerical} parameter search {in} the {current model} on an interval{. The} 
differences {in method} between {the analyses on this manuscript and in the prior researches}~\cite{Fujimoto:2012wv,Fujimoto:2013ki,Fujimoto:2014fka} are summarized:
\begin{itemize}
\item
The relative distances of the point interactions are fixed {by the minimization of the Casimir energy}~\cite{Fujimoto:2017lln}
(see Eq.~(\ref{eq:condition-on-Ls})),
while previously the positions were treated as completely free parameters~\cite{Fujimoto:2012wv,Fujimoto:2013ki,Fujimoto:2014fka}.
\item
Due to {the minimization of the Casimir energy} shown in Eq.~(\ref{eq:MS-Ansatz}),
currently we cannot discuss the flavor mixing in a consistent way.
Here, we take the strategy where at first we focus on only mass hierarchies, which may be justified as the first step to know
how the geometry constrained by the {Casimir energy} works well to reproduce the observed magnitudes of the SM {mass hierarchy},
of course up to the inevitable distortion by mixings.
This attitude will be reasonable especially for the quark sector since the observed mixing angles are small.
On the other hand, we touch effects from mixings by introducing off-diagonal terms by hand phenomenologically
for the charged leptons.
\item
The exact form of the $y$-dependent VEV {of the Case (I) is adopted for the singlet $\Phi$, while an approximate form of the Case (II) solution is used in the prior researches.}
\end{itemize}

Now, the quark and lepton mass matrices are diagonal, where the elements of the matrices are formulated
for $\psi \,(= u,d,e)$ as{
\al{
M^{(\psi)}_{ii} \
&= \widetilde{\mathcal{Y}}_{\psi}\cdot\frac{v_\text{H}}{\sqrt{2}} 
\sqrt{\frac{\widetilde{M}_\Phi^2}{\widetilde{\lambda_\Phi}}}(1-X)^{1/2}\nonumber\\
&\qquad\times
{ \int_{\widetilde{L}_{i-1}}^{\widetilde{L}_{i}} \!\! {d\widetilde{y}} } \ 
\text{sc}\left( \sqrt{\frac{\widetilde{M}_\Phi^{2}(1+X)}{2}} (\widetilde{y} - \widetilde{y}_0), \sqrt{\frac{2X}{1+X}} \right)
\widetilde{g}^{(0)}_{\psi_{i{\rm{L}}}}(\widetilde{y}) \widetilde{f}^{(0)}_{\psi_{i{\rm{R}}}}(\widetilde{y}) \quad
\left( i = 1,2,3 \right),
	\label{eq:Mii_diagonal}
}
where 
\al{
X \equiv \sqrt{ 1 - \frac{4 \widetilde{E}_{\Phi} \widetilde{\lambda}_\Phi}{\widetilde{M}_\Phi^4} },
}
and} the formula is factorized as the product of the flavor independent and dependent parts, and here
all of the parameters {are scaled to the dimensionless ones by the total length of the system $L$, {\it e.g.}, $M_\Phi = \widetilde{M}_\Phi \cdot L^{-1}$} except for the 4d Higgs VEV $v_\text{H} \,(= 246\,\text{GeV})$ {defined as $v_\text{H} \equiv v\sqrt{L}$
from the 5d Higgs VEV $v$}.
We adopted the notation that variables accompanying the tilde symbol are dimensionless, where the normalization part of
the fermion wave functions are made dimensionless following the instruction.
The uniformly located positions of the point interactions are represented as ({\it c.f.} Eq.~(\ref{eq:condition-on-Ls}))
\al{
\widetilde{L}_0 = 0,\quad
\widetilde{L}_1 = \frac{1}{3},\quad
\widetilde{L}_2 = \frac{2}{3},\quad
\widetilde{L}_3 = 1.
	\label{eq:Lwilde-conditions}
}
We comment that the form in Eq.~(\ref{eq:Mii_diagonal}) does not depend on $L$ itself since it describes phenomena of zero modes.
{The} independent parameters for describing the deformation of the flavor dependent part via $\Phi$ are taken as
$X$, $\widetilde{M}_\Phi$, and $\widetilde{y}_0$.
We remind that the five bulk masses of the fermions {${\{\widetilde{M}_{Q}, \widetilde{M}_{\mathcal{U}}, \widetilde{M}_{\mathcal{D}},\widetilde{M}_{L},\widetilde{M}_{\mathcal{E}}\}}$} and the three overall Yukawa couplings ${\{\widetilde{\mathcal{Y}}_{u},\widetilde{\mathcal{Y}}_{d},\widetilde{\mathcal{Y}}_{e}\}}$ also contribute to the
flavor dependent part, where the $10$ parameters {in {Eqs.}~(\ref{para-1}) and (\ref{para-2})} are relevant when we debate with the flavor dependence up to the overall scale
(where one of the Yukawa couplings is included to the overall scale).
The fit observables are the eight mass ratios of the nine SM fermion masses like {$m_\text{others}/m_\text{top}$}.
{We should mention that as was pointed out in the {Ref.}~\cite{Fujimoto:2013ki}, more parameters than the experimental values of the quark/lepton masses do not mean that we can always reproduce the quark and lepton masses in the present scenario. The geometry of the extra dimension tightly restrict the form of the mass matrices to four-zero texture, {see {\it e.g.}, Eq.~(\ref{eq:Mu-example})}. Therefore, it is quite nontrivial whether our model can reproduce the quark and lepton masses
{even though two extra parameters (out of $10$ ones for fitting the eight observables) seem to remain}.}

\subsection{{Results for the numerical research}
\label{sec:diagonal-numerics}}
We found that, even in the current limited setup, when we take the following benchmark,\footnote{
Here, the nearest divergent point of the Jacobi's elliptic function is located at $\widetilde{y} \simeq 1.00249$, which is outside
the region where the interval geometry is elaborated.
}
\al{
\widetilde{M}_\Phi &= 4.3,\quad
\widetilde{y}_0 = 0.33,\quad
X = 0.74, \notag \\
\widetilde{M}_{Q} &= 40,\quad
\widetilde{M}_{\mathcal{U}} = 320,\quad
\widetilde{M}_{\mathcal{D}} = 0.1,\quad
\widetilde{M}_{L} = 100,\quad
\widetilde{M}_{\mathcal{E}} = 0.01,\label{para-1}
}
and take the degrees of freedom of the Yukawa couplings to adjust the third-generation fermions as
\al{
\frac{\widetilde{\mathcal{Y}}_d}{\widetilde{\mathcal{Y}}_u} \simeq 0.124,\quad
\frac{\widetilde{\mathcal{Y}}_e}{\widetilde{\mathcal{Y}}_u} \simeq 0.0454,\label{para-2}
}
the result comes
\al{
\frac{M^{(u)}_{11}}{m_\text{up}} &\simeq 0.746, \quad
\frac{M^{(u)}_{22}}{m_\text{charm}} \simeq 1.08, \quad
\frac{M^{(u)}_{33}}{m_\text{top}} = 1, \notag \\
\frac{M^{(d)}_{11}}{m_\text{down}} &\simeq -1.07, \quad
\frac{M^{(d)}_{22}}{m_\text{strange}} \simeq 0.966, \quad
\frac{M^{(d)}_{33}}{m_\text{bottom}} = 1, \notag \\
\frac{M^{(e)}_{11}}{m_\text{electron}} &\simeq -0.775, \quad
\frac{M^{(e)}_{22}}{m_\text{muon}} \simeq 0.253, \quad
\frac{M^{(e)}_{33}}{m_\text{tauon}} = 1, 
	\label{eq:diagonal-result}
}
where we compare the obtained result with {the derived pole masses} at the leading order from the PDG digits~\cite{Tanabashi:2018oca}.\footnote{
Concrete digits of the central values are as follows:
$m_\text{up} = 2.5\,\text{MeV}$,
$m_\text{down} = 5.2\,\text{MeV}$,
$m_\text{strange} = 110\,\text{MeV}$,
$m_\text{charm} = 1.448\,\text{GeV}$,
$m_\text{bottom} = 4.557\,\text{GeV}$,
$m_\text{top} = 173.0\,\text{GeV}$,
$m_\text{electron} = 0.5109989461\,\text{MeV}$,
$m_\text{muon} = 105.6583745\,\text{MeV}$,
$m_\text{tauon} = 1.77686\,\text{GeV}$, respectively.
}
We reached a good fit for quarks as ratios,\footnote{
On the other hand, these results are sensitive to the modulation of the input parameters in sub-leading magnitudes.
For example, if we change $\widetilde{M}_\Phi$ from $4.3$ to $4.31$,
the accurate fit for quarks turns out to be disrupted as
${M^{(u)}_{11}}/{m_\text{up}} \simeq 0.416$,
${M^{(u)}_{22}}/{m_\text{charm}} \simeq 0.605$,
${M^{(d)}_{11}}/{m_\text{down}} \simeq -0.761$,
${M^{(d)}_{22}}/{m_\text{strange}} \simeq 0.687$,
${M^{(e)}_{11}}/{m_\text{electron}} \simeq -0.504$,
${M^{(e)}_{22}}/{m_\text{muon}} \simeq 0.165$,
respectively, where the other ratios are still unities after suitable adjustments of $\mathcal{Y}_d/\mathcal{Y}_u$ and $\mathcal{Y}_e/\mathcal{Y}_u$.
The position of the nearest divergent point of the elliptic function becomes $\widetilde{y}_0 \simeq 1.00093$.
This point is considered as an inevitable difficulty of the present scenario, where a suitable tuning of inputs is required.
}
while the fit of the charged lepton is {quite} far from being precise.
We do not pay attention to the overall flavor independent part since we may adjust this part by use of $\widetilde{\mathcal{Y}}_u$ and $\widetilde{\lambda}_\Phi$.
The negativity for $M^{(d)}_{11}$, $M^{(e)}_{11}$ originates from the lower tail part of the $\text{sc}$ function,
where now the absolute values of them physically make sense.

\subsection{{{An extra} phenomenological study of the mixing effect {in charged leptons}}
\label{sec:non-diagonal-numerics}}
A possible origin of the {deviated} result in the charged leptons may be that no suitably-enough relevant degree of freedom remains for charged leptons
after we fix the parameters to reproduce the observed mass hierarchy of quarks under the constraints in Eq.~(\ref{eq:Lwilde-conditions}).

{To look into possible improvements of the charged lepton part}, 
we do a phenomenological trial for the charged lepton part, which seems to have a difficulty to regenerate the observed patterns
of their mass eigenvalues.
Now we focus on the speculation that the charged leptons can be mixed largely, because the Pontecorvo--Maki--Nakagawa--Sakata~(PMNS) matrix
which describes the mixings among the left-handed charged leptons and active neutrinos contains large mixing angles.
Under such circumstances mass eigenvalues can alter remarkably by the introduction of sizable non-diagonal terms of mass matrix.
{The existence of non-diagonal elements might be realized when we evaluate the Casimir energy at two-loop order or introduce exotic fermions, which have no {chiral} zero modes and only affect to the Casimir energy. Therefore, to investigate} such a case with {phenomenologically introduced terms} by hand would be
fruitful to know further possibilities of the current scenario{.}

\begin{figure}[t]
\centering
\includegraphics[width=0.60\columnwidth]{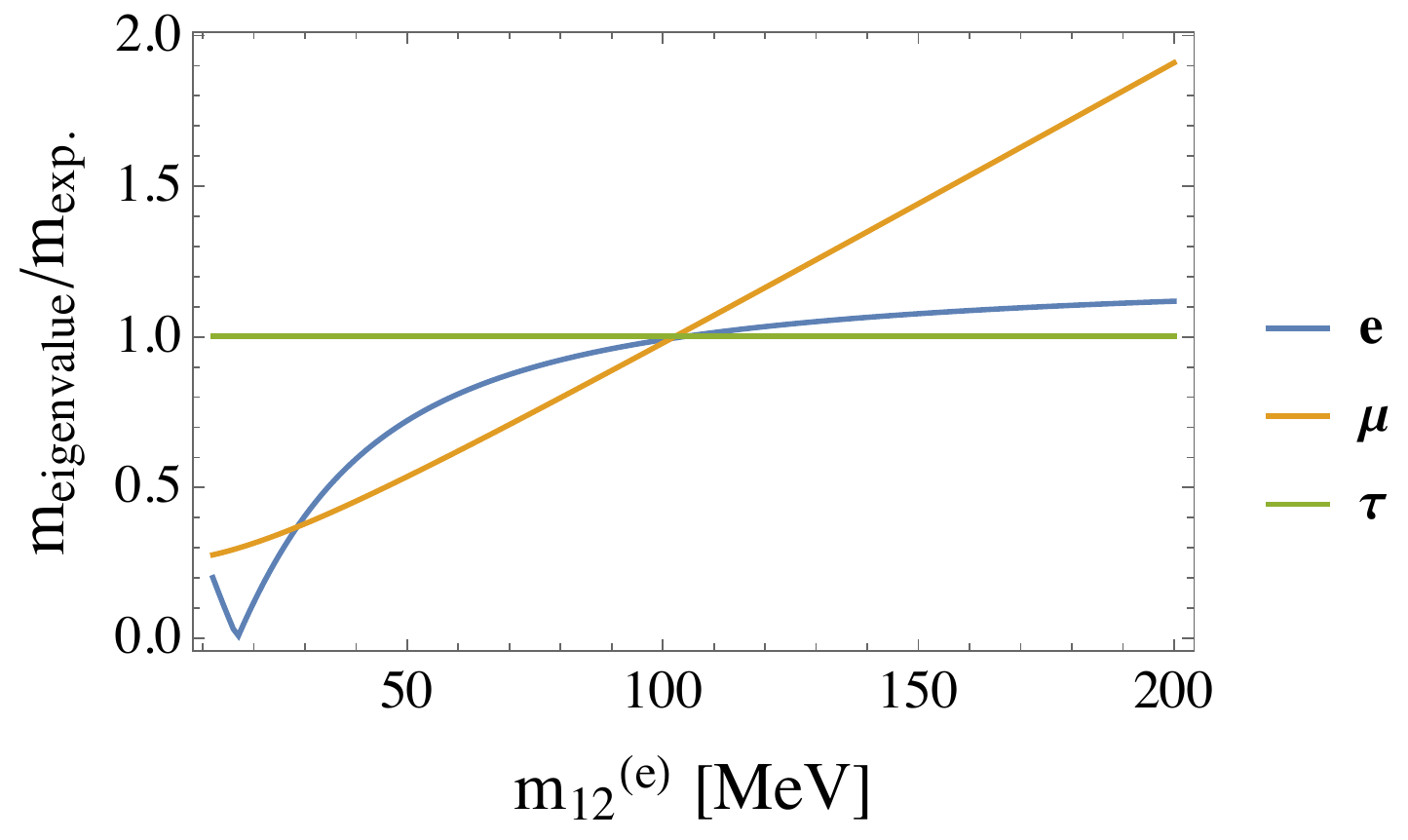}
\caption{
Shifts of mass eigenvalues of the phenomenological charged-lepton mass matrix $\mathcal{M}^{(e)}$ as
functions of $m_{12}^{(e)}$, where $m_{23}^{(e)}$ and $m_{31}^{(e)}$ are pinned down as
$100\,\text{MeV}$ and $11.2\,\text{MeV}$, respectively.
The three eigenvalues get to be very close to the measured digits when $m_{12}^{(e)} \simeq 104\,\text{MeV}$ simultaneously.
}
\label{fig:lepton-correction}
\end{figure}

In what follows, we bring the following phenomenological form for a charged-lepton mass matrix into focus:
\al{
\mathcal{M}^{(e)}
=
\begin{bmatrix}
M^{(e)}_{11} & m_{12}^{(e)} & 0 \\
0 & M^{(e)}_{22} & m_{23}^{(e)} \\
m_{31}^{(e)} & 0 & M^{(e)}_{33}
\end{bmatrix},
}
where we assume that the diagonal inputs take the same digits as in Eq.~(\ref{eq:diagonal-result}) without rounding,
and $m_{12}^{(e)}$, $m_{23}^{(e)}$ and $m_{31}^{(e)}$ are phenomenological parameters.\footnote{
It is pointed out that no $1 \leftrightarrow 3$ mixing term can be realized on an interval.
But it is easy to activate such mixing term after we switch the geometry to $S^1$ (see~\cite{Fujimoto:2013ki}).
}
We found a desirable configuration to reproduce the measurements, where 
$m_{12}^{(e)}$, $m_{23}^{(e)}$ and $m_{31}^{(e)}$ are chosen as 
$104\,\text{MeV}$, $100\,\text{MeV}$ and $11.2\,\text{MeV}$, respectively (see Fig.~\ref{fig:lepton-correction}).
Here, we achieved the good accuracy: 
$(m_\text{eigen}/m_\text{exp.})_\text{electron} \simeq 1.00052$,
$(m_\text{eigen}/m_\text{exp.})_\text{muon} \simeq 1.01612$ and
$(m_\text{eigen}/m_\text{exp.})_\text{tauon} \simeq 1.0016$.
{The} current scope of this issue is just within a naive discussion founded on phenomenological assumptions {so that} it would be worth to investigate the full situation grounded on {higher-loop calculations or modification of the matter contents.}

\section{Conclusion and Discussion
\label{sec:section-5}}
In this paper, we have performed the numerical research in our dynamical model to reproduce the quark/lepton mass hierarchy with the experimental values of the quark/lepton masses. Our model consists of {$10$} point interactions, 
positions of which {were} treated as free parameters in the prior researches~\cite{Fujimoto:2012wv,Fujimoto:2013ki,Fujimoto:2014fka}.
{Their positions have been} determined by the minimization condition of the Casimir energy.
Because of that, the parameters {relevant for fixing the ratios of the fermion masses} have been reduced to {10} from {20} in our model and only the remnant {10} parameters are tools for reemergence {of} the {eight} experimental values of the quark and lepton mass ratios. As {was} pointed out in the {Ref.}~\cite{Fujimoto:2013ki}, more parameters than the experimental values of the quark/lepton masses do not mean that we can always reproduce the quark and lepton masses in our model. The geometry of the extra dimension tightly restricts the form of the mass matrices to {four-zero texture, see Eq.~(\ref{eq:Mu-example})}. Therefore, it is quite nontrivial whether our model can reproduce the quark and lepton masses.

To obtain the desired mass hierarchy, we introduced the extra-dimension coordinate-dependent VEV of the gauge singlet scalar {$\langle\Phi(y)\rangle$, which is also obtained dynamically by minimizing the potential of the scalar $\Phi$
(refer to Eq.~(\ref{eq:S_singlet}))}. 
As a result of {our} numerical analysis, we found that there is a parameter set, 
{which can reproduce the experimental values of the quark well, while some difficulties remain for the charged leptons.
On the other hand, we pointed out that the mixing terms among the charged leptons will resolve remaining discrepancies,
even though it is not justified fully in the theoretical point of view within the current scope.}


{A definite next step is to contemplate how to realize non-diagonal terms in the quarks and the charged leptons,
which are necessary ingredients to generate the observed textures of the mixings represented by the CKM and PMNS matrices.}
The calculation of the Casimir energy was achieved at one-loop order, and consequently 
the positions of the point interactions are determined as to divide the interval extra dimension equally among three, 
{\it i.e.} the mass matrices of quarks and leptons automatically become diagonal {ones}.
If off-diagonal components appear after introducing a suitable mechanism{,} 
the eigenvalues of the mass matrices deviate from the diagonal components of the mass matrices and the experimental values might be recovered with {flavor mixings}, as is discussed in subsection~\ref{sec:non-diagonal-numerics}.

One {possible} idea to generate off-diagonal components in the mass matrices is to introduce exotic fermions. 
Boundary conditions for the exotic fermions should be chosen to produce 
no massless chiral zero modes, and then they will contribute only to 
the Casimir energy at low energies. A possible boundary condition 
for the exotic fermion $\Psi^{\textrm{ex}}(x,y)$ is given by
%
\begin{align}
	\Psi^{\rm ex}_{\rm L}(x,y)=0\quad \text{at}\quad y=0,L_{1}+\varepsilon,\qquad \Psi^{\textrm{ex}}_{\rm R}(x,y)=0\quad \text{at}\quad y=L_{1}-\varepsilon, L.
	\label{eq5.1}
\end{align}
%
It is noted that the exotic fermion feels the point interaction 
only at $y=L_{1}$ but not the one at $y=L_{2}$. 
The boundary condition (\ref{eq5.1}) can prohibit the presence 
of chiral zero modes \cite{Fujimoto:2011kf}, and will force the point 
interaction at $y=L_{1}$ to move from $L_{1}=L/3$ to $L_{1}>L/3$. 
This is because the exotic fermion contributes to the position $y=L_{1}$ 
of the point interaction as to divide the interval extra dimension equally among two.
After combining the contributions of the exotic fermion with the (5d)
SM fermions, the minimization condition of the Casimir energy is 
modified and the off-diagonal component $M_{12}$ (and probably other 
off-diagonal components) can appear due to $L_{1}>L/3$. 
It should be mentioned that there is a way to obtain $L_{1}<L/3$ 
by extending the mechanism. We can introduce extra three point interactions 
only for the exotic fermion and impose the BCs at $y=L'_{2}, L'_{3}, L'_{4}$ 
in addition to $y=L_{1}$. In this case, the exotic fermion may contribute 
to the position $y=L_{1}$ of the point interaction as to divide 
the interval extra dimension equally among four, in which $L_{1}<L/3$ 
can be realized from the modified minimization condition of the Casimir energy. 
The key ingredient is that this mechanism accomplishes the appearance 
of off-diagonal components dynamically and the contribution of the exotic 
fermion might be the same order as the (5d) SM fermions, so that 
large off-diagonal components could emerge. 
If bulk masses of some exotic fermions are chosen to be much larger than 
those of the (5d) SM fermions, their effects on the Casimir energy are 
expected to be limited and will give small contribution to off-diagonal 
components of the the mass matrices.
Therefore we can expect that both of the small/large mixing in the 
quark/lepton sector might be obtained dynamically by use of this mechanism.

Another idea is 
{to take account of higher-order effects} of the Casimir energy. 
Since the flavor {structures} of the quarks and leptons are nontrivial, the effects of the matter contents at two-loop order might produce
{corrections} to the minimization conditions of the Casimir energy. 
That may {lead to} off-diagonal components of the mass matrices through the {modulations} of the positions of the point interactions. Origin of modulations may also be the (zero-mode)-(KK-mode) mixing which can make small mass perturbations.
The issues stated above remain to be pursued as the future work.


\section*{Acknowledgements}
This work was supported by JSPS KAKENHI Grant Number JP 18K03649~(Y.F., M.S. and K.T.).
K.N. has been supported by the European Union through the European Regional Development Fund -- the Competitiveness and 
Cohesion Operational Programme (KK.01.1.1.06),
and the grant funded from the European Structural and Investment Funds,
RBI-TWINN-SIN. The authors thank T. Inoue, S. Sato, Y. Hashimoto and T. Yamamoto for useful discussions and also thank T. Miura for discussions in the early stage of this work.

\bibliographystyle{utphys}
\bibliography{references,Flavor_May2019,references-extra,references-extra-2,additional-refs}

\end{document}